%% file: moriond.tex
\documentclass[11pt]{article}
\pdfoutput=1
\usepackage{moriond,epsfig}
\usepackage{xspace}
\usepackage{subfig}
\input{macros}

\begin{document}
%\vspace*{4cm}
\title{MEASUREMENTS OF TOP PROPERTIES AT THE TEVATRON}

\author{U. HUSEMANN on behalf of the CDF and D\O{} Collaborations}

\address{Department of Physics, Yale University, P.O. Box 208120, New Haven, CT 06520--8120, USA}

\maketitle
\abstracts{\vspace{-5mm} The large data samples of thousands of top
  events collected at the Tevatron experiments CDF and D\O{} allow for
  a variety of measurements to analyze the properties of the top
  quark. Guided by the question ``Is the top quark observed at the
  Tevatron really the top quark of the standard model,'' we present
  Tevatron analyses studying the top production mechanism including
  resonant $\ttbar$ production, the \VminusA structure of the $t\to
  Wb$ decay vertex, the charge of the top quark, and single-top
  production via flavor-changing neutral currents.}
{\vspace{-5mm} \small {\it Keywords}: Hadron Collider, Tevatron, Heavy Quark Production, Top Quark Properties}

%\vspace{-105mm}
%\begin{flushright}
%CDF/PUB/TOP/PUBLIC/8813
%\end{flushright}
%\vspace{97mm}

\section{Introduction}

At the Tevatron collider at Fermi National Accelerator Laboratory,
protons and antiprotons are collided at a center-of-mass energy of
1.96\tev. The Tevatron provides the highest energies currently
available at a hadron collider and is the only collider with
sufficient energy to produce top quarks. The two multi-purpose
experiments at the Tevatron, CDF II~\cite{Acosta:2004yw} and
D\O~\cite{Abazov:2005pn}, have studied production and decays of top
quarks in great detail. The focus of this article will be on
measurements of top properties at CDF and D\O{} (other than mass and
cross section). The event selection for top properties measurements is
built on the experience gained in the mass and cross section analyses
in the ``lepton+jets'' channel ($\ttbar\to Wb\,Wb\to \ell\nu b\,qqb$)
and the ``dilepton'' channel ($\ttbar\to Wb\,Wb\to \ell\nu b\,\ell\nu
b$). Also the background composition, with $W$ production in
association with jets as the dominant source, and systematic
uncertainties, mainly coming from determining the jet energy
scale, are similar to those in the top mass and cross section
analyses.

\section{Top Pair Production}

\subsection{Fraction of \ttbar Pairs Produced via Gluon-Gluon Fusion}

In the standard model (SM), the production of \ttbar pairs at the
Tevatron is dominated by the process of \qqbar annihilation. The
contribution of the $gg$ fusion channel to the \ttbar production cross
section amounts to $15\pm 5\%$, where the large uncertainty is due to
poor knowledge of the gluon parton distribution function inside the
proton~\cite{Cacciari:2003fi,Kidonakis:2003qe}. The CDF collaboration
has developed two complementary methods to measure the fraction of
\ttbar pairs produced via $gg$ fusion. With datasets of up to 1\invfb
of integrated luminosity, both methods are dominated by statistical
uncertainties and are therefore expected to improve with more data.

\begin{figure}[t]
  \centering
  \subfloat[]{\label{fig:gg_rutgers}\includegraphics[height=52mm]{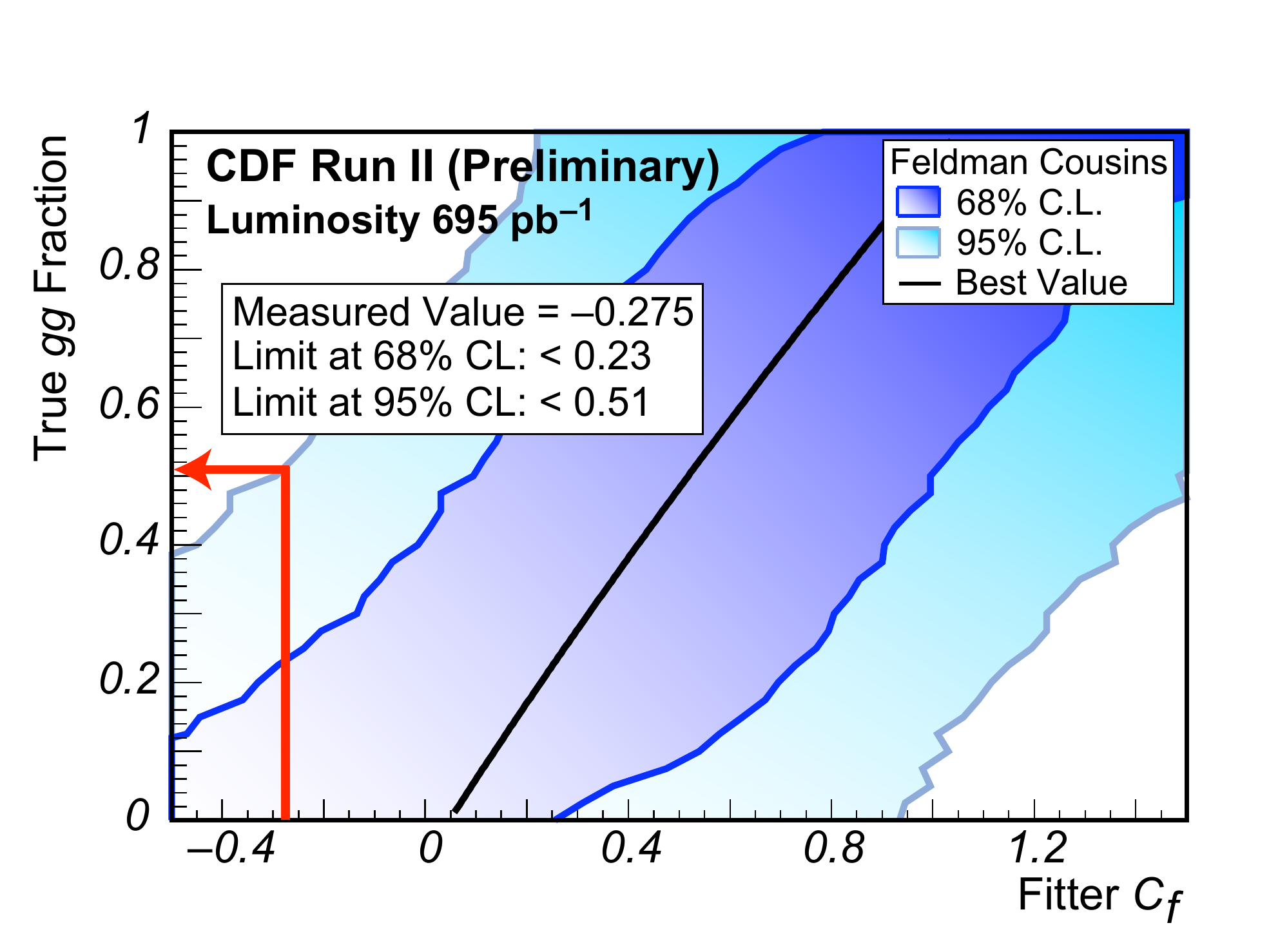}}
  \hspace{10mm}
  \subfloat[]{\label{fig:gg_toronto}\includegraphics[height=48mm]{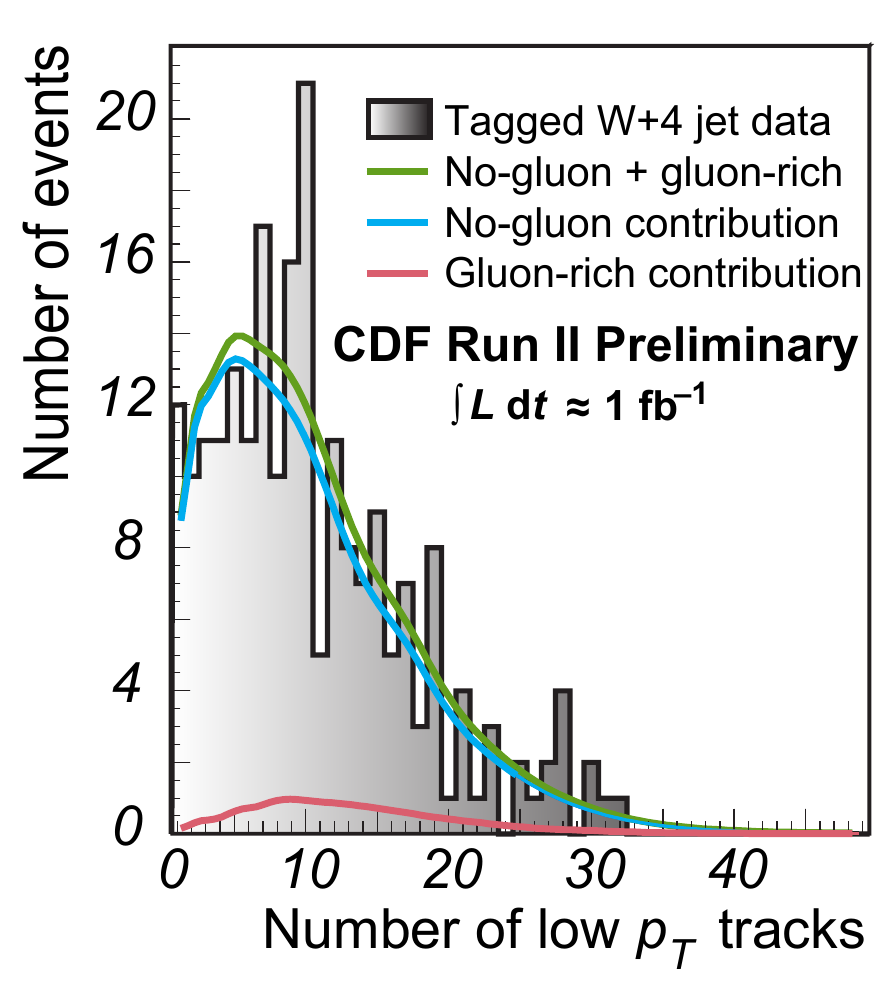}}
  \caption{Measurements of the $gg$ fraction in \ttbar
    production. (a)~Feldman-Cousins band obtained for neural network
    discriminant method. (b)~Distribution of low-\pt tracks fitted by
    gluon-rich and no-gluon templates in $W$+4-jet data.}
\end{figure}

One method~\cite{ggfraction_Rutgers} utilizes an artificial neural
network (NN) to distinguish the processes $\qqbar\to\ttbar$,
$gg\to\ttbar$, and $\qqbar/gg\to W$+jets. The NN input comprises
the velocity and angle of the top quark, and the angles of the
three decay products in the off-diagonal spin basis. From the resulting
NN discriminant, an upper limit on the $gg$ fraction is derived via a
Feldman-Cousins method that includes systematic uncertainties. As
shown in Fig.~\ref{fig:gg_rutgers}, the measurement yields a limit of
%\[
%\frac{\sigma(gg\to\ttbar)}{\sigma(\ppbar\to\ttbar)} < 0.51\quad(95\%\:\mathrm{C.L.}).
%\]
${\sigma(gg\to\ttbar)}/{\sigma(\ppbar\to\ttbar)} < 0.51$ at 95\% C.L.

The other method~\cite{ggfraction_Toronto} is based on the observation
that the number of tracks with small transverse momenta \pt in the range of
0.3--2.9\gevc is strongly correlated with the number of gluons in the
event. The correlation is calibrated using gluon-rich and gluon-free
control samples in the data, and then extrapolated to the top-rich
sample of $W$+4~jets. Gluon-rich and no-gluon templates are fitted
to the distribution of low-\pt tracks in the data, as depicted in
Fig.~\ref{fig:gg_toronto}, to extract a $gg$ fraction of
%\[
%\frac{\sigma(gg\to\ttbar)}{\sigma(\ppbar\to\ttbar)}=0.01\statsyst{0.16}{0.07}.
%\]
${\sigma(gg\to\ttbar)}/{\sigma(\ppbar\to\ttbar)}=0.01\statsyst{0.16}{0.07}$.

\subsection{Resonant Production of \ttbar Pairs}
Several extensions of the SM predict \ttbar production
from the decay of heavy particles. In one particular
model~\cite{Hill:1993hs}, the heavy particle is assumed to be a narrow
$Z'$ resonance that couples strongly only to third generation quarks
and does not couple to leptons (``leptophobic $Z'$'').

\begin{figure}[t]
  \centering
  \subfloat[]{\label{fig:ttbar_resonance}\includegraphics[height=58mm]{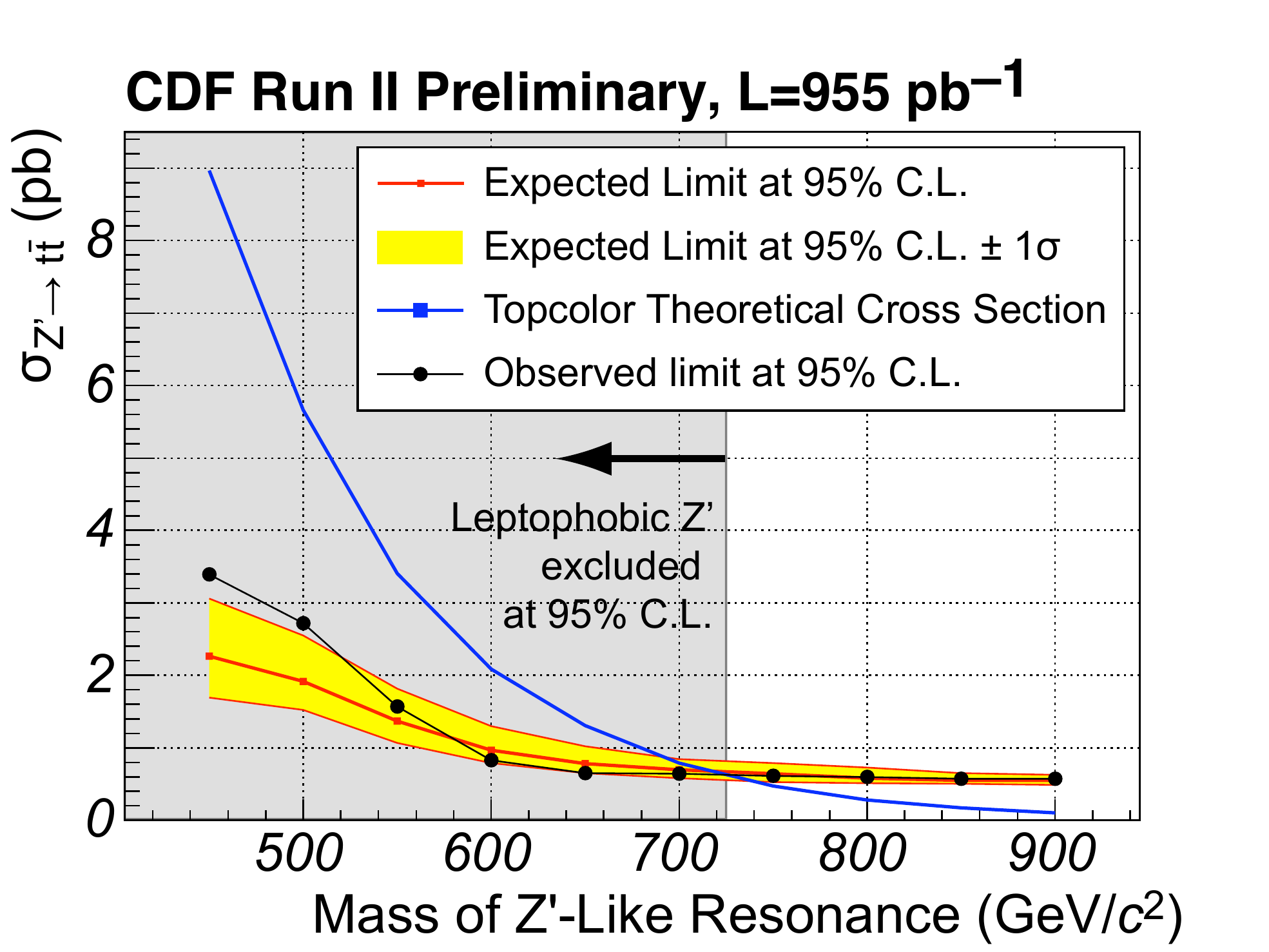}}
  \hspace{10mm}
  \subfloat[]{\label{fig:whel_summary}\includegraphics[height=58mm]{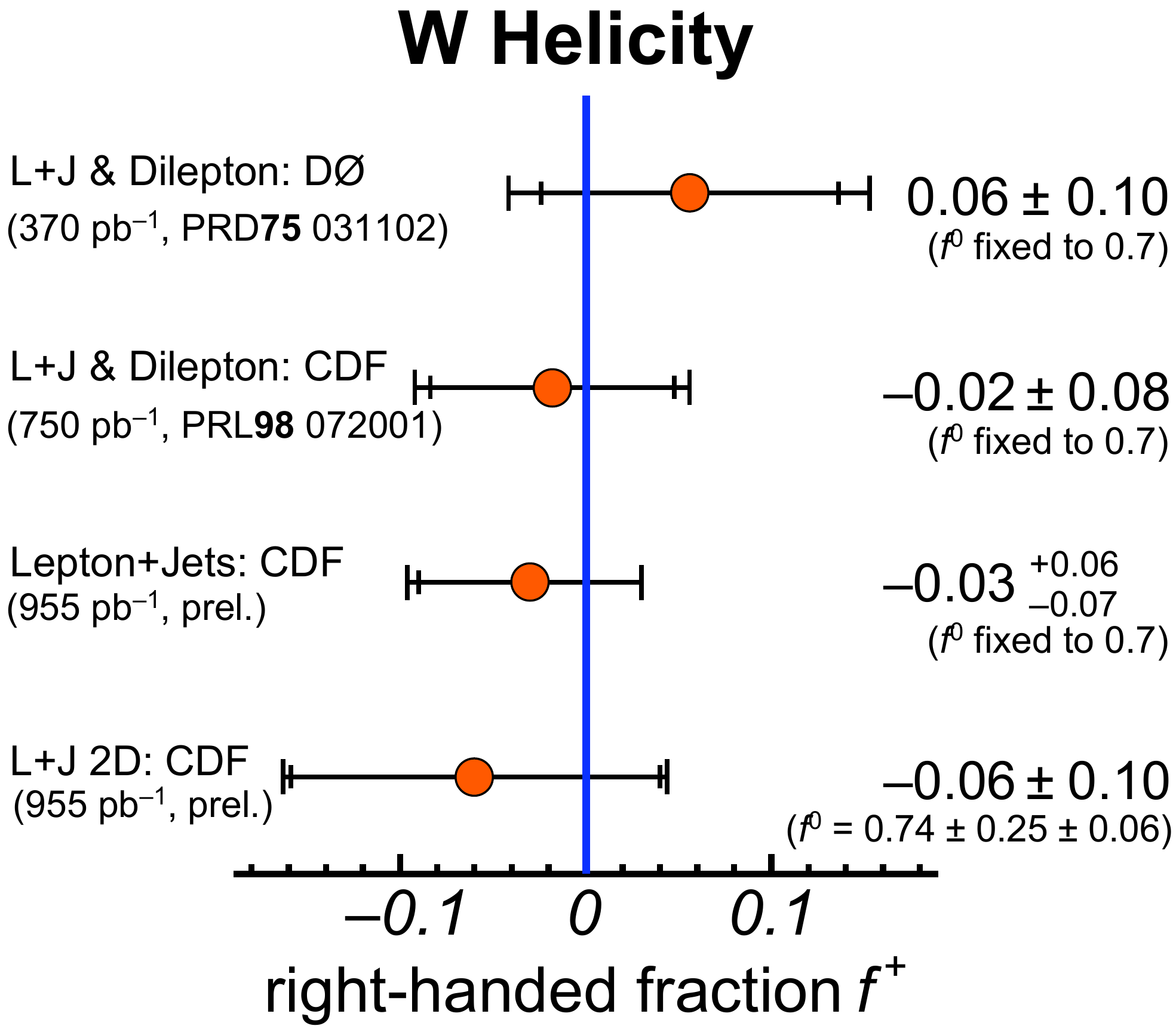}}
  \caption{(a)~Limit on the production cross section of leptophobic
    $Z'$ bosons. The shaded area is excluded at 95\% C.L. (b)~Summary
    of $W$ helicity measurements. Shown are fits to the right-handed
    helicity fraction $f^+$ in different analyses.}
\end{figure}

The CDF collaboration has studied the production of a $Z'$-like heavy
\ttbar resonance~\cite{ttbar_resonance}. The \ttbar invariant mass
\Mtt is reconstructed with the help of a kinematic fitter developed
for top mass measurements. The \Mtt distribution in the data is
compared to the expectation for SM \ttbar production and non-\ttbar
background. The background includes $W+$jets events, QCD multijet
events, and diboson production ($WW$, $WZ$, $ZZ$), and was estimated
with a method that combines input from Monte Carlo simulations and data
control samples. The data are compatible with SM \ttbar production, so
that a limit on the $Z'$ production cross section can be derived,
see Fig.~\ref{fig:ttbar_resonance}. From the comparison with the cross
section prediction for leptophobic $Z'$, CDF obtains a limit of
$M_{Z'} > 725\gevcsq$ at 95\% C.L.
%\[
%M_{Z'} > 725\gevcsq \quad(95\%\:\mathrm{C.L.}).
%\]

\section{Helicity of \boldmath$W$\unboldmath{} Bosons from Top Decays}

Due to its small lifetime of less than $10^{-24}\unit{s}$, the top
quark decays before it hadronizes. As a consequence, the complete spin
information is transferred to its daughter particles. In the SM, top
quarks decay to a $W$ boson and a $b$ quark almost exclusively.  The
spin-1 $W$ boson has three possible helicity states: longitudinal,
left-handed, and right-handed. The \VminusA structure of the $tWb$
vertex in the SM does not allow for a right-handed state, so that any
sizable admixture of right-handed $W$'s would be a sign of new
physics. The longitudinal fraction of the $W$ helicity in the SM is
determined by the Yukawa coupling of the top quark and amounts to
$f^0\approx 0.7$.

Both CDF and D\O{} have measured the right-handed helicity fraction
$f^+$ of $W$ bosons from top
decays~\cite{Abazov:2006hb,Abulencia:2006iy,Whel_karlsruhe,Whel_FNAL}. The
analyses use different techniques to reconstruct the observable
\costh, the angle between the top boost direction and the charged
lepton in the $W$ rest frame, or the correlated quantity $M_{\ell b}$,
the invariant mass of the charged lepton and the $b$ jet. Due to the
limited size of the data samples, most
analyses~\cite{Abazov:2006hb,Abulencia:2006iy,Whel_karlsruhe} check
the consistency of $f^0$ with the SM, assuming $f^+=0$, and fix the
value to $f^0=0.7$ before $f^+$ is measured. With 1\invfb of data, a
simultaneous fit to $f^0$ and $f^+$ becomes
feasible~\cite{Whel_FNAL}. As shown in Fig.~\ref{fig:whel_summary},
all measurements are compatible with the SM prediction of $f^+=0$.

\section{Top Charge}
In the SM, the top quark and the bottom quark form a left-handed
isospin doublet with charges $(2/3\,e,-1/3\,e)_L$, where $e$ is the
elementary charge. Fits to electroweak precision data can be improved
using a theoretical model with a top mass of $270\gevcsq$ and an
exotic right-handed quark doublet with charges $(-1/3\,e,-4/3\,e)_R$ that
mixes with right-handed $b$ quarks~\cite{Chang:1998pt}. In this model,
the exotic replacement for the top quark has a charge of $-4/3\,e$.

The CDF collaboration has performed a measurement to test if the top
charge is $2/3\,e$ or $-4/3\,e$ ~\cite{topcharge}. The measurement
consists of three steps. The $W$ charge is obtained via the charge of
the lepton in the decay $W\to \ell\nu$. To reconstruct the top from
the decay $t\to Wb$, the $W$ is paired with a $b$ jet, using a
kinematic fitter (lepton+jets channel) or a cut on $M_{\ell b}$
(dilepton channel). Finally, the observable ``jet charge,'' a weighted
sum of the charge of all tracks that form a jet, is employed to
measure the flavor of the $b$ jet. The product of $W$ charge and jet
charge is used to distinguish the exotic model from the SM, see
Fig.~\ref{fig:topcharge}.

CDF has tested the consistency of the data with the SM and the exotic
model with a hypothesis test, with the null hypothesis that the SM is
correct. If the exotic model were correct, 81\% of all measurements
would return $p$-values below 0.01 (probability to incorrectly reject
the SM, chosen \emph{a priori}). The measured $p$-value of 0.35 shows that
the data are consistent with the SM, and the exotic model is excluded
at 81\% C.L.

\section{Single-Top Production via Flavor-Changing Neutral Currents}

Flavor-changing neutral currents (FCNC) in the top quark sector are
heavily suppressed in the SM. For example, the branching fraction
$\mathcal{B}(t\to gc)$ is approximately $5\times10^{-12}$, far below
the reach of present and future hadron collider
experiments~\cite{Aguilar-Saavedra:2004wm}.  Any FCNC signal at the
Tevatron would be a sign of physics beyond the SM.

\begin{figure}[t]
  \centering
  \subfloat[]{\label{fig:topcharge}\includegraphics[height=55mm]{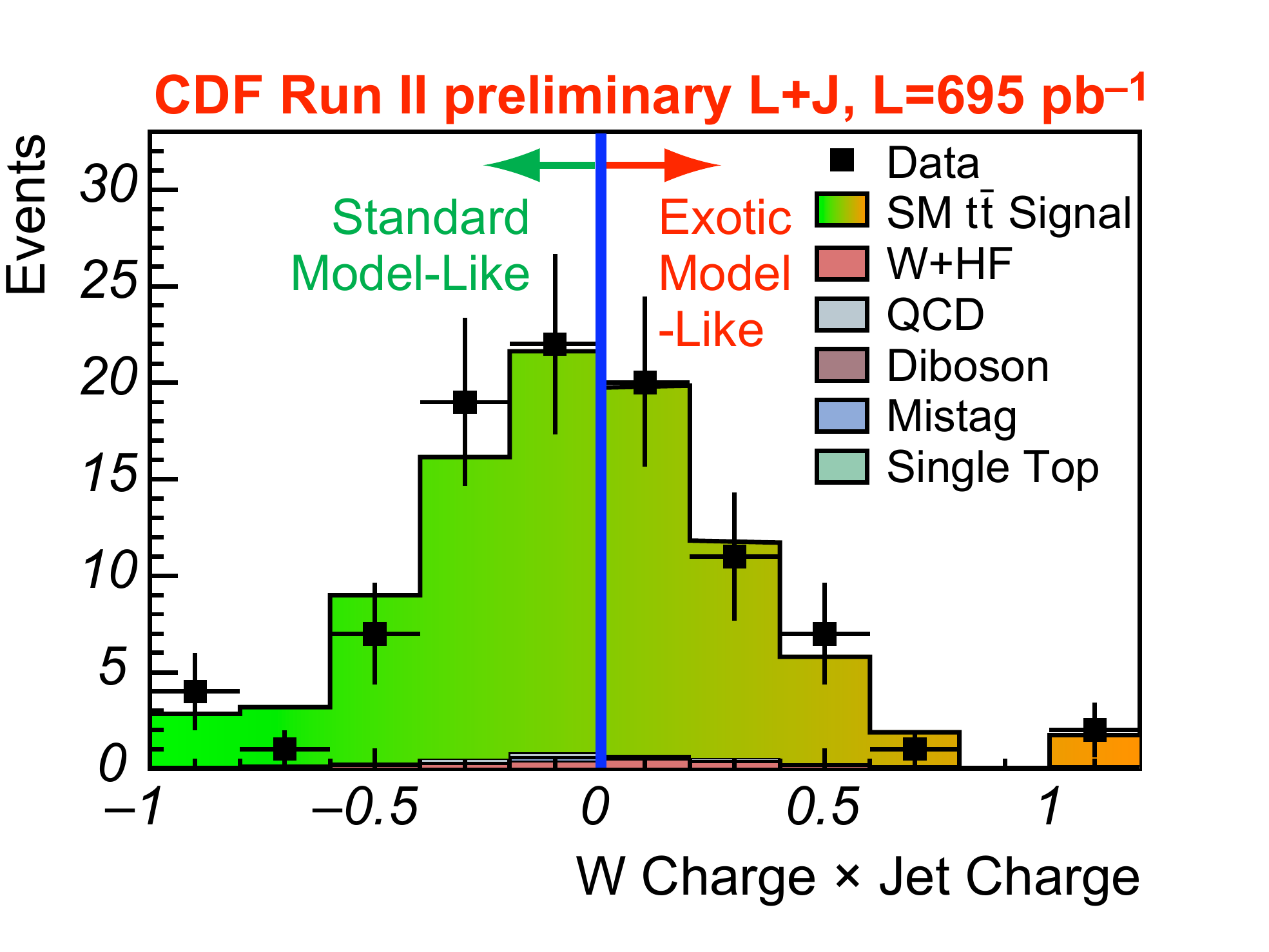}}
  \hspace{5mm}
  \subfloat[]{\label{fig:fcnc}\includegraphics[height=48mm]{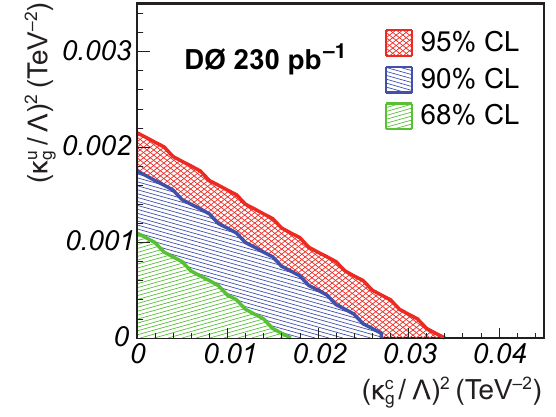}}
  \caption{(a) Top charge measurement via jet charge. The data are
    compared to the predictions of the SM for \ttbar signal and
    background. (b) Exclusion contours for the couplings
    $(\kappa/\Lambda)^2$ of the flavor-changing neutral current
    vertices $tgu$ and $tgc$.}
\end{figure}

The D\O{} collaboration has searched for FCNC in the
production of single top quarks~\cite{Abazov:2007ev}. In the presence
of the processes $u g \to t$ and $c g \to t$, the single top
production rate is enhanced. D\O{} has deployed a NN to discriminate
the kinematics of the FCNC signal from \ttbar, $W$+jets, and QCD
multijet backgrounds. The data agrees very well with the SM
prediction, so that an upper limit on the FCNC couplings $tgu$ and
$tgc$ could be derived. Fig.~\ref{fig:fcnc} shows the upper limit as a
function of the strengths $(\kappa/\Lambda)^2$ of the $tgu$ and $tgc$
coupling, where $\kappa$ is the coupling parameter in the Lagrangian,
and $\Lambda$ is a generic new physics scale. The D\O{} measurement
improves upon previous measurements, prominently by the HERA
experiments, by a factor of 3--11.

\section*{Acknowledgments}
It is a pleasure to thank the organizers of Moriond QCD for creating a
very relaxing and inspiring atmosphere at the workshop.  This
material is based upon work supported by the U.S.\ Department of
Energy under Award Number DE-FG02-92ER40704.

\section*{References}

%\nocite{*}

%\bibliographystyle{myamsplain}
\bibliographystyle{unsrt}
\bibliography{moriond}

\end{document}

%% file: macros.tex
\newcommand{\qbar}{\ensuremath{\overline{q}}\xspace}

\newcommand{\tbar}{\ensuremath{\overline{t}\xspace}}

\newcommand{\pbar}{\ensuremath{\overline{p}\xspace}}

\newcommand{\ttbar}{\ensuremath{{t\tbar}}\xspace}
\newcommand{\qqbar}{\ensuremath{{q\qbar}}\xspace}
\newcommand{\ppbar}{\ensuremath{{p\pbar}}\xspace}

\newcommand{\costh}{\ensuremath{\cos\theta^*}\xspace}
\newcommand{\pt}{\ensuremath{p_{T}}\xspace}

\newcommand{\Mtt}{\ensuremath{M_{\ttbar}}\xspace}

\newcommand{\unit}[1]{\ensuremath{\:\mathrm{#1}}\xspace}
\newcommand{\statsyst}[2]{\ensuremath{\pm#1\mathrm{(stat.)}\pm#2\mathrm{(syst.)}\xspace}}

\newcommand{\gevc}{\ensuremath{\unit{GeV}/c}\xspace}
\newcommand{\gevcsq}{\ensuremath{\unit{GeV}/c^{2}}\xspace}
\newcommand{\tev}{\unit{TeV}}
\newcommand{\invfb}{\unit{fb^{-1}}}

\newcommand{\VminusA}{\ensuremath{V\!-\!A}\xspace}